\documentclass[12pt]{article}
\usepackage{amsmath}
\usepackage{amssymb}
\usepackage{array}
\usepackage{graphicx}
\usepackage{latexsym}
\usepackage{textcomp}

\newtheorem{theorem}{Theorem}[section]
\newtheorem{proposition}{Proposition}[section]

\newtheorem{lemma}[theorem]{Lemma}

\newtheorem{definition}{Definition}[section]

\begin{document}

\title{Evolutionary games under incompetence \thanks{This research is funded by the Australian Research Council Discovery Grant DP160101236.}
}
%\subtitle{A new approach to adaptation dynamics}

\author{Maria Kleshnina \footnote{Centre for Applications in Natural Resource Mathematics (CARM), School of Mathematics and Physics, University of Queensland, St Lucia, Queensland 4072, Australia}, Jerzy A. Filar $^{\dagger}$, \\ 
Vladimir Ejov \footnote{School of Computer Science, Engineering \& Mathematics, Flinders University, Adelaide, South Australia 5001, Australia}, Jody C. McKerral \footnote{School of Biological Sciences, Flinders University, Adelaide, South Australia 5001, Australia}
}

% \institute{M. Kleshnina \at
%               Centre for Applications in Natural Resource Mathematics (CARM), School of Mathematics and Physics, University of Queensland, St Lucia, Queensland 4072, Australia\\
%               \email{m.kleshnina@uq.edu.au} 
%            \and
%            J.A. Filar	\at
%               Centre for Applications in Natural Resource Mathematics (CARM), School of Mathematics and Physics, University of Queensland, St Lucia, Queensland 4072, Australia\\
%              \email{j.filar@uq.edu.au}     
%            \and
%            V. Ejov	\at
%               School of Computer Science, Engineering \& Mathematics, Flinders University, Adelaide, South Australia 5001, Australia\\
%               \email{v.ejov@flinders.edu.au}    
%            \and
%            J.C. McKerral	\at
%               School of Biological Sciences, Flinders University, Adelaide, South Australia 5001, Australia\\
%               School of Computer Science, Engineering \& Mathematics, Flinders University, Adelaide, South Australia 5001, Australia\\
%               \email{jody.fisher@flinders.edu.au} 
% }

%\date{Received: date / Accepted: date}
% The correct dates will be entered by the editor

\maketitle

\begin{abstract} The adaptation process of a species to a new environment is a significant area of study in biology. As part of natural selection, adaptation is a mutation process which improves survival skills and reproductive functions of species. Here, we investigate this process by combining the idea of incompetence with evolutionary game theory. In the sense of evolution, incompetence and training can be interpreted as a special learning process. With focus on the social side of the problem, we analyze the influence of incompetence on behavior of species. We introduce an incompetence parameter into a learning function in a single-population game and analyze its effect on the outcome of the replicator dynamics. Incompetence can change the outcome of the game and its dynamics, indicating its significance within what are inherently imperfect natural systems.

%\keywords{Evolutionary game theory \and Incompetence \and Matrix games \and Replicator dynamics}
% \PACS{PACS code1 \and PACS code2 \and more}
% \subclass{MSC code1 \and MSC code2 \and more}
\end{abstract}

\section{Introduction}
\label{intro}
Evolutionary game theory, first introduced in 1973 by Maynard Smith and Price, analyses interactions between different populations of animals or their offspring during their lives, and has since become an area of ongoing research interest \cite{Smith1973,Bomze1983,Hofbauer2003,Zeeman1980}. It aims to answer the important ecological question of which population (or strategy) is the most stable from invasion by mutants. As a part of natural selection, adaptation can be a mutation process which improves survival skills and reproductive functions of species. This paper aims to provide a novel approach to adaptation processes in new environments from the social point of view. This is achieved by introducing the notion of incompetence, whereby players may be imperfect in executing their strategies. Hence evolutionary dynamics are considered under the assumption that species improve their level of competence according to a prescribed learning scheme.

A classical assumption in evolutionary game theory is that any player in a game chooses a strategy from a strategy set and executes it with probability one. However, this assumption may be overly simplistic for a realistic model, because players might not be experts in their preferred actions. For example, if a fight between a young and an aged lion unfolds, we could not predict the real outcome of the game as the experienced lion may be more skillful and less prone to errors than the young one, even if lacking in youthful vigor. This corresponds to potential mistakes in executing strategies. Mathematically, this means that the probability of executing a chosen strategy would be less than one. 

In the context of evolutionary games, the idea of allowing players to make errors is not entirely new. It has been described in many different ways: as a mutation process alone \cite{Hadeler1981,Stadler1992}, a process of language learning \cite{Komarova2004,Komarova2001,Nowak2001}, different experimental learning processes \cite{FudenbergLevine,Hopkins2002,McKelvey1995,Selten1991}, adaptation dynamics \cite{Levin2003}, and as environmental noise \cite{Binmore1995,Mao2002}. In addition, the unpredictability of some model aspects (such as behavioral or environmental ``noise'') may be approximated by taking into consideration the perturbations of different parameters. This approximation is also related to the idea of players who make small errors, called ``trembling hands'' \cite{Selten1975}, and as a result of these effects, stability of the equilibria is affected. An attempt to generalize perturbations caused by players' mistakes, in the sense of behavioral errors in normal form games, was made by Beck et al. in their paper {\itshape Incompetence and impact of training in bimatrix games} \cite{Filar2012}, and in Beck's PhD thesis {\itshape Incompetence, training and changing capabilities in game theory} \cite{Beck2013} for matrix and bimatrix stationary games. 

In parallel to evolutionary game theory, the dynamical systems perspective is frequently adopted via the analysis of the associated replicator equations \cite{Taylor1978}. The behavior of the replicator dynamics depends on initial conditions, parameters' values and the structure of the payoff matrix, and have been extensively studied \cite{Bomze1983,Bomze1986,Nowak2006,Zeeman1980}. However, there is no existing research examining the behavior of replicator dynamics under incompetence.

\subsection{Mathematical Background for Replicator Dynamics}

There are many approaches for establishing long run scenarios for evolutionary games. A fundamental set of replicator equations was proposed by Taylor and Jonker in \cite{Taylor1978}. Consider the fitness matrix $R=[r_{ij}]$, $i,j=1, \ldots, n$. Let $\mathbf{x}=(x_i)$ denote the frequency of strategy $i$. Then the expected payoff (or fitness) of strategy $i$ is defined by the formula
\begin{equation}
\label{Fitnessi}
	f_i=\sum_{j=1}^{n} x_jr_{ij}=e^T_i R x=(Rx)_i,
\end{equation}
where $e_i$ is the $i$-th vector in the unit basis. The average fitness payoff of the population is then defined by the scalar
\begin{equation}
\label{MeanFitness}
	\phi=\sum_{i=1}^{n} x_if_i=x^T R x.
\end{equation}

The appealing replicator equation capturing the dynamics of the strategy $i$ is then postulated to be
\begin{equation*}
	\dot{x}_i=x_i(f_i-\phi),\;i=1,...,n, 
\end{equation*}
or in matrix form
\begin{equation}
\label{ReplicatorDynamics}
	\dot{x}_i=x_i\left( \left(Rx \right)_i-x^T R x \right),\;i=1,...,n.
\end{equation}

\subsection{Mathematical Background to Incompetence}

To incorporate incompetence, let us now consider a matrix game with a $n \times n$ payoff matrix $R$. In a classical sense, each player chooses an action (a pure strategy) and that choice results in a deterministic payoff: there is an underlying assumption that players are able to execute the actions that they have chosen. However, when incompetence is introduced this is no longer the case \cite{Beck2013}.

Incompetence is incorporated into the game by assigning probabilities that the actions selected by players will not coincide with the executed actions. The set of all such probabilities determines the incompetence matrix for each player. Let $q(A_j | A_i)$ be the probability that player 1 executes action $A_j$ given that he selects action $A_i$. Obviously, for any chosen action the probabilities of all possible executed actions sum to $1$, that is
$\sum_{j=1}^{n} q(A_j | A_i)=1,\; \forall i=1,...,n, $
and
$0\leq q(A_j | A_i) \leq 1,\; \forall i,j.$
The set of all these probabilities for player 1 makes up his incompetence matrix $Q$:
\begin{equation}
\label{IncompetenceMatrix}
Q=\left(\begin{array}{cccc}
q(A_1 | A_1)& q(A_2 | A_1)& ...& q(A_n | A_1)\\
q(A_1 | A_2)& q(A_2 | A_2)& ...& q(A_n | A_2)\\
\vdots& \vdots& \ddots& \vdots\\
q(A_1 | A_n)& q(A_2 | A_n)& ...& q(A_n | A_n)
\end{array} \right).
\end{equation}
Note that $Q$ is a stochastic matrix. We assume that the incompetence matrix for the second player is a transpose matrix of the first player, as both of them are from the same population.

Now, we can construct a matrix game with incompetence. Suppose that the two players independently select actions $i$ and $k$, then the probability that this results in executed actions $j$ and $h$ respectively is
\begin{equation*}
p(A_j , A_h | A_i, A_k)=q(A_j | A_i)q(A_h | A_k). 
\end{equation*}

In order to simplify notation, instead of $q(A_j | A_i)$, we will now use the notation $q_{ij}$. The expected reward can therefore be determined as a function of the selectable strategies as
\begin{equation}
\label{IncompReward}
r^Q_{ik}=\sum_{j=1}^{n} \sum_{h=1}^{n} p(A_j , A_h | A_i, A_k) r_{jh}=\sum_{j=1}^{n} \sum_{h=1}^{n} q_{ij} q_{kh} r_{jh}, 
\end{equation}

and the values $r^Q_{ik}$ define the entries of the payoff matrix $R^Q$.

The matrix form of (\ref{IncompetenceMatrix}) gives the relationship 

\begin{equation}
\label{IncompetenceRM}
R^Q=Q R Q^T. 
\end{equation}

To define players' incompetence, we introduce an incompetence parameter $\lambda\in [0,1]$ which measures the training progress of players on the trajectory from the ``starting'' level of incompetence, $S$, to the ``final'' level of incompetence, $F$. These trajectories can be of any type. In order to simplify the analysis we shall initially consider a linear trajectory such as:
\begin{equation}
\label{incomp}
Q(\lambda)=(1-\lambda) S+\lambda F,\; \lambda\in[0,1].
\end{equation}

Here both $S$ and $F$ are also stochastic matrices. In some cases, $F$ will be an identity matrix indicating the final level of full competence. We use simply $R(\lambda)$ to notate a new incompetent fitness matrix from (\ref{IncompetenceRM}). Note that, $R(0)=SRS^T$, $R(1)=FRF^T$, and hence $R(1)=R$ if $F=I$.

The key questions here are to discover the rules of how to adapt in the most efficient way, and determine if there are any critical points, $\lambda^c$, on the adaptation trajectory of the population. Such critical points may determine the adequate level of adaptation, after which species do not have to improve their survival skills. 

In the next section we introduce the idea of incompetence in the classical replicator dynamics. In Section \ref{sec:2} we demonstrate how this adaptation process affects the selection outcome of the classical Hawk-Dove game. Then, in Section \ref{sec:3} we provide results that can be used when analyzing evolutionary games under incompetence to examine the influence of incompetence on the outcome of the replicator dynamics. Finally, we demonstrate our results on the extended Hawk-Dove-Retaliator game in Section \ref{sec:4}.

\section{Evolutionary dynamics under incompetence}
\label{sec:1}

Let us now consider how probabilities of mistakes may describe interaction errors in an evolutionary context. At first, one can imagine that a particular population of species is immersed into a new environment; this could be described by a natural or anthropogenic migration process. Assume that there are only a finite number, $n$, of available behavioral strategies for each species. Two individuals interact by choosing and executing strategies, and achieve payoffs, defined by the $n\times n$ matrix $R$. Next, we also assume that each individual may choose any action from the available set of strategies. We assume that in the new habitat they make errors, executing different strategies from the ones that they chose, with probabilities $q_{ij}$, where $i$ and $j$ are varying from $1$ to $n$. The set of all conditional probabilities $q_{ij}$ for the population makes up its incompetence matrix $Q$ from (\ref{IncompetenceMatrix}).

Thus, whenever two individuals interact, they may both change their behavior from one strategy to another during the execution of the play. Assume that both opponents are able to determine each other's selected strategy. This is a natural assumption as species are able to ``recognise'' other species' behavior by their body language. However, even if the opponent has chosen one particular strategy, there is a chance for them to execute another according to the experience or competence in the strategy choice. Hence, each species from the population achieves an expected payoff for chosen strategy $i$ whenever their opponent chooses strategy $k$ according to (\ref{IncompReward}) or, in simplified notation

\begin{equation}
\label{IncompetentReward}
r_{ik}(\lambda)=\sum_{j=1}^{n} \sum_{h=1}^{n} q_{ij}(\lambda)q_{hk}(\lambda) r_{jh}, \;i,k=1,..,n,
\end{equation}
where $q_{ij}(\lambda)$ denotes the $(i,j)^{\text{th}}$ entry of $Q(\lambda)$.

\subsection{Replicator dynamics under incompetence} 

Now we can introduce the adaptation dynamics into the evolution of the population. Here, the adaptation process of a population to a new environment can be constructed as interactions between the environment's individuals, which over time reduce their probabilities of making errors, that is, as $\lambda$ changes from $0$ to $1$. Then, according to (\ref{Fitnessi})-(\ref{ReplicatorDynamics}), for a new matrix game under incompetence given by $R(\lambda)$, one may write down the following equations of the expected fitness for strategy $i$ 
\begin{equation}
\label{IncompFitness}
f_i(\lambda)=\sum_{j=1}^{n} r_{ij}(\lambda)x_j=e_i^T R(\lambda) x,
\end{equation}
and for the mean fitness payoff of the population 
\begin{equation}
\label{IncompMeanFitness}
	\phi(\lambda)=\sum_{i=1}^{n} x_if_i(\lambda)=x^T R(\lambda) x.
\end{equation}

Hence, the replicator dynamics are given by 
\begin{equation*}
	\dot{x}_i=x_i(f_i(\lambda)-\phi(\lambda)),\;i=1,...,n, 
\end{equation*}
or in a matrix form
\begin{equation}
\label{IncomReplicatorDynamics}
	\dot{x}_i=x_i((R(\lambda)x)_i-x^T R(\lambda) x).
\end{equation}
Thus we obtain a dynamic system (\ref{IncomReplicatorDynamics}), where $R(\lambda)$ is quadratic in $\lambda$. An important feature of these systems is that the time scale of replicator dynamics for $x(t)$ might not coincide with the time scale of adaptation dynamics for $\lambda$. This means that individuals may study the environment much faster or much slower than they reproduce. Indeed, we begin our analysis of the dynamics of the system with $\lambda$ fixed, with the goal of discovering the underlying effect of the incompetence parameter. 

In a strict sense, the new system given by (\ref{IncomReplicatorDynamics}) is a perturbed evolutionary game, and perturbations depend on the parameter $\lambda$. As $\lambda$ tends to $1$ the game under incompetence gets closer to the original game given by $R$. 

\section{Motivating example: a Hawk-Dove game}
\label{sec:2}

In his book \cite{Smith1982}, Maynard Smith analyzed the basic example of evolutionary game theory which is called ``The Hawk-Dove game''. In this game, the interaction between two bird types, of Hawk (H) and Dove (D), is observed. In a generic interaction, these birds need to divide some resource of value $b$, for example, territory or food. If they are both from the Hawk population, then they fight and lose some resource of value $c$, perhaps representing a cost of injury, and divide the rest of the resource in half. If they are both Dove, then they simply divide the resource without any losses. If Hawk and Dove interact, then Hawk captures the resource and Dove receives nothing. The payoff matrix here is given by

$$R=\left( \begin{array}{ccc}
\frac{b-c}{2}& &b\\
\\
0& &\frac{b}{2}\\
\end{array} \right).$$ 

The first row and column correspond to the Hawk strategy while the second row and column denote the Dove strategy. It is a well-studied example and it was shown that the result depends on the structure of the payoff matrix. In particular, if the price of injury is high, then stable coexistence is possible. Otherwise, we obtain the situation when the aggressive strategy dominates the passive one.

One particular example to illustrate this game is an interaction between a naturally aggressive person and a passive one. It can be easily imagined that the passive person is inclined to be scared and might run away when he or she becomes a victim of aggression. Obviously, a fight in response to the aggression is not the action that is expected from aggressive. On the other hand, the aggressive person is expected to fight no matter what the circumstances are. However, we might observe that in some cases the passive person could fight back with extra aggression, whereas, the aggressive person can become frightened and run away. This behavioral unpredictability may be more likely when players are in a new environment with which they are unfamiliar. 

Next, we construct an incompetent case of the Hawk-Dove game with the reward matrix $R(\lambda)$ calculated from (\ref{IncompetenceRM})-(\ref{incomp}). Consider the case when stable coexistence of two strategies is possible with $b=2$ and $c=4$. Then,

$$R=\left( \begin{array}{ccc}
-1& &2\\
0& &1\\
\end{array} \right).$$ 

Here Dove and Hawk stably coexist and there exists a unique stable frequency of Hawks, $x_H$, which is given by $x_H=\frac{b}{c}=\frac{1}{2}.$

The game is described via one payoff matrix for both players: for the first player the matrix is $R$, and for the second player the matrix is $R^T$, and if we assume that players from the same population are equally incompetent, then only one matrix $(Q,Q^T)$ measures players' level of incompetence. Consider the starting incompetence level $S$ as 
$$S=\left( \begin{array}{ccc}
0.3& &0.7\\
0.6& &0.4\\
\end{array} \right),$$ 
and obtain $Q(\lambda)$ from (\ref{IncompetenceMatrix})
$$Q(\lambda)=\left( \begin{array}{ccc}
0.7\lambda+0.3& &0.7-0.7\lambda\\
0.6-0.6\lambda& &0.6\lambda+0.4\\
\end{array} \right).$$ 

We use a simplification of the reward matrix with $0$ on the diagonal from
\begin{equation}
\label{SimplifiedRM}
\tilde{R}(\lambda)=R(\lambda)-\mathbf{d}_{R(\lambda)}\mathbf{u}^T,
\end{equation}
where $\mathbf{d}_{R(\lambda)}$ is a column-vector consisting of the diagonal elements of $R(\lambda)$, $\textbf{u}$ is a column-vector of ones. This formula helps to simplify replicator dynamics as it is known that this transformation does not affect the dynamics \cite{Zeeman1980}. The reward matrix $\tilde{R}(\lambda)$ then can be written as
$$\tilde{R}(\lambda)=\left( \begin{array}{ccc}
0& &1.56\lambda^2-0.62\lambda+0.06\\
1.82\lambda^2-0.94\lambda+0.12& &0\\
\end{array} \right).$$

It can be easily shown that there are three possible situations depending on the value of $\lambda$. The first is if Hawk and Dove stably coexist: thus, there exists a stable equilibrium $$x_H=\frac{6\lambda-1}{13\lambda-3}$$ and this is the case for $\lambda\in [0,\frac{1}{6})\cup (\frac{2}{7},1]$.  We observe an interesting result when species achieve the level of incompetence  $\lambda=\frac{1}{6}$. Hawk become extinct and we obtain a population consisting of all Dove. Hence, we enter the new interval $\lambda\in [\frac{1}{6},\frac{3}{13}]$ where we obtain a population consisting of all-Doves, as this strategy dominates. However, mistakes give Hawks a hope for rebirth as the probability a Dove acting like a Hawk is
$$q_{21}=0.6-0.6\lambda.$$
For example, for $\lambda=0.2$, a Dove will revive a Hawk strategy in $48\%$ cases. As players learn more, $\lambda$ falls into the interval $(\frac{3}{13},\frac{2}{7}]$ and a Hawk strategy becomes an ESS, meaning that it is preferable and the frequency of Hawks starts to grow rapidly. At $\lambda=\frac{2}{7}$ we expect to obtain a population consisting of all-Hawks. This is a reverse situation to $\lambda=\frac{1}{6}$, where Hawk acts like Dove in $50\%$ cases, which preserves the latter from extinction. As species keep learning, the system falls into the interval of $\lambda > \frac{2}{7}$ where Hawk and Dove stably coexist.

\section{Perturbations under incompetence}
\label{sec:3}

In this section we provide results that give tips on understanding of the game dynamics when incompetence is established. As has been illustrated in the Hawk-Dove example, there exist transition points on the adaptation trajectory that change qualitative characteristics of the system. Let us define these points as critical values of the incompetence parameter:

\begin{definition}
{\em A critical value of the incompetence parameter}, $\lambda^c$, is the bifurcation point of the replicator dynamics. Then, let $\Lambda$ be the set of all such critical values of $\lambda$.
\end{definition}

Bifurcation points occur in dynamics as singular points of the Jacobian matrix. However, let us also define the set of values of the incompetence parameter when the determinant of the reward matrix $R$ equals to zero, that is

\begin{definition}
Let $Z=\left\{\lambda\in [0,1]\; |\; \det(\tilde{R}(\lambda))=0\right\}$. We call it {\em the set of singular points} of the incompetent game $\tilde{R}(\lambda)$.
\end{definition}

By the definition of a bifurcation, if the system is stable, then in the intervals between $\lambda^c_1$ and $\lambda^c_2$ the fixed points preserve their qualitative behavior.
That is, in order to understand where the game experiences transitions, we want to find these critical values of the incompetence parameter. Strictly speaking, fixed points of the incompetent replicator dynamics depend on the incompetence parameter, that is $\mathbf{\tilde{x}}(\lambda^c)$, however, we will use the simplified notation $\mathbf{\tilde{x}}$.

Let us first analyze a fixed point at the vertex $i$ of the simplex $S^n$. According to \cite{Bomze1986} the eigenvalues of the Jacobian for such points are $0$ and ${\tilde{r}_{ji}:j\neq i}$. Hence, when changes in the incompetence parameter cause changes in the sign of the corresponding elements of the payoff matrix $\tilde{R}(\lambda)$ we observe changes in stability of such points.

Next consider a fixed point on the edge, i.e. the point $\mathbf{x}=\alpha \mathbf{e}_i+(1-\alpha) \mathbf{e}_k,$ where $\mathbf{e}_i$ and $\mathbf{e}_k$ are the unit basis vectors. That is, at this fixed point we obtain only two survived strategies. Then, we can easily show that

\begin{lemma}
If: 

(a) $\mathbf{\tilde{x}}=\alpha\mathbf{e}_i+(1-\alpha)\mathbf{e}_j,$ where $\alpha\in (0,1)$, is a stable fixed point, and 

(b) $\lambda^c$ is a value where $\phi_k(\lambda)=\mathbf{e}_k\tilde{R}(\lambda)\mathbf{\tilde{x}}-\mathbf{\tilde{x}}\tilde{R}(\lambda)\mathbf{\tilde{x}},$ for some $k\neq i,j$, changes sign from $<0$ to $>0$,

then $\mathbf{\tilde{x}}$ changes its qualitative behavior at $\lambda^c$ and becomes unstable.
\end{lemma}

{\itshape Proof.} 
According to Bomze's result on the fixed points $\mathbf{\tilde{x}}=\alpha\mathbf{e}_i+(1-\alpha)\mathbf{e}_j,$ \cite{Bomze1986}, the values 
$$\phi_k(\lambda)=\mathbf{e}_k\tilde{R}(\lambda)\mathbf{\tilde{x}}-\mathbf{\tilde{x}}\tilde{R}(\lambda)\mathbf{\tilde{x}},k\neq i,j,$$
are the eigenvalues of the Jacobian at $\mathbf{\tilde{x}}$. 

The stability of the fixed point implies that all eigenvalues of the Jacobian have negative real parts. Hence, the changes in sign of at least one eigenvalue will lead to the changes in qualitative behavior of the fixed point.

\hfill $\square$

\textbf{Remark:} we should note that Lemma 1 holds for unstable fixed points as well. However, for these points to change their qualitative behavior to the stable fixed point all $\phi_k(\lambda),\forall k\neq i,j$ have to become negative.

Next, it is well-known that if $\tilde{\mathbf{x}}$ is a fixed point of the replicator dynamics, then 
\begin{equation}
\label{JacobianFP}
(J_{\tilde{x}}\tilde{\mathbf{x}})=-\phi(\tilde{x})\tilde{\mathbf{x}},
\end{equation}
where $J_{\tilde{x}}$ is a Jacobian matrix evaluated at the fixed point $\tilde{\mathbf{x}}$ \cite{Bomze1986}. We use this fact in order to continue our analysis.

Let us first notice that we are using a simplified version of the reward matrix from (\ref{SimplifiedRM}). Next, construct the $i$-th replicator equation in terms of elements of the $n\times n$ reward matrix $\tilde{R}$ and a vector $\mathbf{x}$ for $k,j=1,\ldots,n$:
\begin{equation}
\label{RepDynExt}
\dot{x}_i=x_i \left(\sum_{j\neq i} x_j r_{ij}- \left( \sum_{j\neq i} x_i x_j(r_{ij}+r_{ji})+\sum_{l\neq i} \sum_{k\neq i,l} x_l x_k r_{kl} \right) \right).
\end{equation}

Consider the general case of the $i$-th component of $J_x \mathbf{x}$ for some vector $\mathbf{x}$:
\begin{equation}
\label{JacobianX}
(J_x x)_i=x_i \left(2\sum_{j\neq i} x_j r_{ij} - \left( 3\sum_{j\neq i} x_i x_j (r_{ij}+r_{ji}) + 3\sum_{l\neq i}\sum_{k\neq i,l} x_l x_k r_{kl} \right) \right).
\end{equation}
Comparing equations (\ref{RepDynExt}) and (\ref{JacobianX}) it follows
$$(Jx)_i= 2x_i(f_i-\phi)-x_i \phi =2\dot{x}_i-x_i\phi.$$
Hence, if $\tilde{\mathbf{x}}$ is a fixed point, then $\dot{x}_i=0,\forall i,$ and we obtain the required result.

Next, let us analyze the behavior of the incompetent replicator dynamics at the fixed point $\tilde{\mathbf{x}}$ such that
$$\dot{x}_i=\tilde{x}_i((\tilde{R}(\lambda)\tilde{x})_i-\tilde{x}^T\tilde{R}(\lambda)\tilde{x})=0,$$
and in the matrix form we obtain
$$\tilde{X}\tilde{R}(\lambda)\tilde{\mathbf{x}}=\tilde{X}(\tilde{\mathbf{x}}^T\tilde{R}(\lambda)\tilde{\mathbf{x}})=\tilde{X}\phi(\tilde{\mathbf{x}})\textbf{1}=\phi(\tilde{x})\tilde{\mathbf{x}},$$
where $\tilde{X}$ is the diagonal matrix with $\tilde{x}_i$ on the diagonal. From (\ref{JacobianFP}) we obtain
\begin{equation}
\label{RDandJacobian}
\tilde{X}\tilde{R}(\lambda)\tilde{\mathbf{x}}=-J_{\tilde{x}}\tilde{\mathbf{x}}=\phi(\tilde{x})\tilde{\mathbf{x}}.
\end{equation}

Further, let us define $\lambda^c$ for which bifurcations of the replicator dynamics occur when the mean fitness of the population $\tilde{\mathbf{x}}$ equals zero in the following way.

\begin{definition}
Let $\lambda^c$ be {\em a balanced bifurcation parameter value} of the fixed point $\tilde{\mathbf{x}}$ when the mean fitness $\phi(\tilde{\mathbf{x}},\lambda^c)=0$.
\end{definition}

We can now formulate the next result.

\begin{lemma}
If $\mathbf{\tilde{x}}$ is an interior fixed point, i.e. $\tilde{x}_i>0,\forall i$, then every balanced bifurcation parameter value, $\lambda^c$, is also a singular point of $\tilde{R}(\lambda)$.
\end{lemma}

{\itshape Proof.} 
Because $\phi(\tilde{\mathbf{x}},\lambda^c)=0$ and $\tilde{\mathbf{x}}>\mathbf{0}$, equation (\ref{RDandJacobian}) implies that $\tilde{X}\tilde{R}(\lambda^c)$ is singular and hence 
$$ \det(\tilde{X}\tilde{R}(\lambda^c))=\prod_{i=1}^n x_i \times \det(\tilde{R}(\lambda^c))=0.$$

\hfill $\square$

Having a fixed point in the interior of the simplex is a nice property, but it is a rare case. For example, games might possess fixed points on the boundaries of the simplex or one might even observe heteroclinic cycles. However, using a result by Taylor and Jonker, which says that for ESS the fitness of extinct strategies is less than the mean fitness of ESS \cite{Taylor1978}, we can formulate the following proposition.

\begin{proposition}
(i) If $\mathbf{\tilde{x}}=(\tilde{x}_1,\ldots,\tilde{x}_{n-1},0)$ is an ESS, then every balanced bifurcation parameter value, $\lambda^c$, is also a singular point of $\tilde{R}(\lambda)$.

(ii) If $\mathbf{\tilde{x}}=(\tilde{x}_1,\ldots,\tilde{x}_{n-1},0)$ is not an ESS and $\lambda^c$ is a balanced bifurcation parameter value, then $\tilde{x}_j(\lambda^c)=\frac{1}{n-1},\forall j\neq n$.
\end{proposition}

{\itshape Proof of part (i).} 
From (\ref{RDandJacobian}) we obtain
\begin{equation*}
    \begin{matrix}
    [\tilde{R}(\lambda^c)\tilde{\mathbf{x}}]_i & =
    & \left\{
    \begin{matrix}
    0,& \forall i\neq n\\
    f_n,& i=n.\\
    \end{matrix} \right.
    \end{matrix} 
\end{equation*}

Suppose $\tilde{R}(\lambda^c)^{-1}$ exists. Then
$$\tilde{R}(\lambda^c)\tilde{\mathbf{x}}=f_n\mathbf{e}_n$$
and
$$\tilde{\mathbf{x}}=f_n \tilde{R}^{-1}(\lambda^c)\mathbf{e}_n.$$

Then, by Cramer's rule,
$$x_j=\frac{|\tilde{R}_j(\lambda^c)|}{|\tilde{R}(\lambda^c)|},$$
where $\tilde{R}_j(\lambda^c)$ is the same as $\tilde{R}(\lambda^c)$ except that $j$-th column is $f_n\mathbf{e}_n$.

As $x_n=0$ we obtain $|\tilde{R}_n(\lambda^c)|=(-1)^{2n}f_n|\bar{R}_{nn}(\lambda^c)|=0,$ where $\bar{R}_{nn}(\lambda^c)$ is a corresponding co-factor. Then, 
\begin{equation}
\label{detCofR=0}
|\bar{R}_{nn}(\lambda^c)|=0.
\end{equation}

Also, $\forall j\neq n$ we obtain $\tilde{x}_j(\lambda^c)>0$ and 
\begin{equation}
\label{xjComponent}
    \begin{matrix}
    \tilde{x}_j(\lambda^c)=\frac{1}{|\tilde{R}(\lambda^c)|} & \times
      & \left|
    \begin{matrix}
    0&r_{12}&\ldots&0&\ldots&r_{1,n-1}&r_{1,n}\\
    \vdots&\vdots&\ddots&\vdots&\ddots&\vdots&\vdots\\
    r_{n,1}&r_{n,2}&\ldots&f_n&\ldots&r_{n,n-1}&0\\
    \end{matrix} \right| & =
    \frac{(-1)^{n+j}}{|\tilde{R}(\lambda^c)|} |\bar{R}_{n,j}(\lambda^c)|. & \\
    \end{matrix} 
\end{equation}

Also, the determinant of the reward matrix expanded by the last row gives
\begin{equation}
\label{detCofR}
|\tilde{R}(\lambda^c)|=\sum_{j=1}^{n-1} (-1)^{n+j} r_{n,j} |\bar{R}_{n,j}(\lambda^c)|.
\end{equation}

Then, from (\ref{xjComponent}) for each $j\neq n$, we obtain
$$(-1)^{n+j}|\bar{R}_{n,j}(\lambda^c)|=x_j|\tilde{R}(\lambda^c)|.$$

Substitute into (\ref{detCofR}) 
$$|\tilde{R}(\lambda^c)|=\left( \sum_{j=1}^{n-1} r_{n,j} x_j \right) |\tilde{R}(\lambda^c)|,$$
which gives
\begin{equation}
\label{fitness1}
1=\sum_{j=1}^{n-1} r_{n,j} x_j =f_n.
\end{equation}
Equation (\ref{fitness1}) is a contradiction as $\tilde{\mathbf{x}}$ is an ESS and for $x_n=0$ we have $f_n\leq \phi(\tilde{\mathbf{x}})=0$.

{\itshape Proof of part (ii).} 
Furthermore, we can show that if $\mathbf{\tilde{x}}=(x_1,...,x_{n-1},0)$ is not an ESS and $|\tilde{R}(\lambda^c)|\neq 0$, then $f_n=1$ and the $\tilde{\mathbf{x}}$ point lies in the ``center'' of the facet $x_n=0.$
That is, from (\ref{fitness1}) we know that $f_n=1,$ and from (\ref{xjComp2}) that for each $j\neq n$ we have
\begin{equation}
\label{xjComp2}
0<x_j=\frac{(-1)^{n+j}}{|\tilde{R}(\lambda^c)|} |\bar{R}_{n,j}(\lambda^c)|=\frac{r_{n,j}}{\sum_{j=1}^{n-1}r_{n,j}}=\frac{1}{\theta}r_{n,j},
\end{equation}
as $\sum_{j=1}^{n-1}x_j=1.$ Substitute (\ref{xjComp2}) in (\ref{fitness1}) to obtain
\begin{equation*}
1=\frac{\sum_{j=1}^{n-1}r^2_{n,j}}{\sum_{j=1}^{n-1}r_{n,j}},
\end{equation*}
and hence 
\begin{equation}
\label{summoverr}
\sum_{j=1}^{n-1}r^2_{n,j}=\sum_{j=1}^{n-1}r_{n,j}.
\end{equation}
Since all $r_{n,j}$ have the same sign by (\ref{xjComp2}), (\ref{summoverr}) implies that $r_{n,j}>0,j=1,\ldots,n-1.$ Then, it also implies $r_{n,j}=1$ for $j=1,\ldots,n-1$ and 
\begin{equation}
\label{xj}
x_j=\frac{1}{n-1},\;j=1,\ldots,n-1.
\end{equation}

\hfill $\square$

We considered the extreme cases with an interior equilibrium, with one strategy becoming extinct and with only one or two strategies surviving. Consideration of other cases is planned for the future research. However, these results give an easy way to verify if there exists a crucial transition point on the adaptation trajectory for games with ESS, especially for low-dimensional systems, that is, $n\leq 3$. In order to continue our analysis, we recall a definition from \cite{Zeeman1980}. 

\begin{definition}
A property of the dynamic system is called {\em robust} if it is preserved under small perturbations.
\end{definition}

Once we know the nature of bifurcation points, we observe that the number of transitions in the game is finite in games with $n\leq 3$. Let us define a value of $\lambda\in (0,1)$ after which no transition is possible where we arrive in the case with sufficiently small perturbations under incompetence. 

\begin{definition}
If there exists $\delta>0$ such that $|| Q(\lambda)-I ||\leq \delta$, where $\delta$ depends on $\lambda^u$ and $\lambda^u=\max \lambda^c$ is the maximal critical value of the incompetence parameter for a fixed point $\mathbf{\tilde{x}}$, then we shall call such perturbations under incompetence {\em sufficiently small perturbations} for this point.
\end{definition}

In other words, if we know that there exists $\lambda^u\in [0,1]$ sufficiently close to $1$, then in the interval $(\lambda^u,1]$ no bifurcations of the fixed point occur and the game preserves its robust properties. Hence, if the population is familiar enough with the environment, species are more likely to behave as in the original game, and the game tends to preserve the same behavioral habits as the original one. Furthermore, we can formulate the next result.

\begin{theorem}
If the game $\tilde{R}$ possesses an ESS, $\tilde{x}$, and $|| Q(\lambda)-I ||\leq \delta(\lambda^u)$, then the incompetent game $\tilde{R}(\lambda)$, when $\lambda\in (\lambda^u,1]$, possesses an ESS, $\tilde{x}(\lambda)$, and 
\begin{equation}
\label{ESSlim}
\lim_{\lambda\rightarrow 1^{-}} \tilde{x}(\lambda) = \tilde{x}.
\end{equation}
\end{theorem}

{\itshape Proof.} 

We know that the evolutionary stability of the game implies local stability and resistance to small perturbations \cite{Bomze1986,Huttegger2008,Huttegger2010}. 

The fixed point $\tilde{\mathbf{x}}(\lambda)$ does not experience any bifurcations when $\lambda>\lambda^u$. Next, we know that all real parts of eigenvalues at this point preserve their sign. At $\lambda=1$ we obtain $\tilde{\mathbf{x}}(\lambda)=\tilde{\mathbf{x}}$, that is, all eigenvalues of $\tilde{\mathbf{x}}$ have negative real parts, hence we obtain that all eigenvalues of $\tilde{\mathbf{x}}(\lambda)$ have negative real parts. This implies that $\tilde{\mathbf{x}}(\lambda)$ is hyperbolic, and, hence, the incompetent replicator dynamics is locally structurally stable.

From the previous statement we obtain that the initial replicator dynamics and incompetent replicator dynamics for $\lambda>\lambda^u$ are locally topologically equivalent as perturbations under incompetence are smooth and points are both hyperbolic. Then $\tilde{\mathbf{x}}(\lambda)$ is an ESS for $\tilde{R}(\lambda)$ and hence by \cite{Bomze1986}, for $\lambda>\lambda^u$

\begin{equation*}
||\tilde{\mathbf{x}}(\lambda)-\tilde{\mathbf{x}}||<\epsilon,\;\;\forall \epsilon>0.
\end{equation*}

\hfill $\square$

We should notice that all results obtained in this section describe the evolution of the strategy choice. However, according to the incompetence matrix $Q(\lambda)$ for any given $\lambda$ and strategy choice, $\mathbf{\tilde{x}}(\lambda)$, we observe a stochastic behavior of the species, $\mathbf{\tilde{y}}(\lambda)$, affected by their incompetence as a result of

\begin{equation}
\label{execution}
\mathbf{\tilde{y}}(\lambda)=Q(\lambda) \mathbf{\tilde{x}}(\lambda).
\end{equation}

\section{Three Strategies: Hawks, Doves and Retaliators (HDR)} 
\label{sec:4}

Let us next demonstrate our results on the $3$-dimensional extension of the Hawk-Dove game. Imagine that we add a new type of species called {\itshape ``Retaliators''} to the system. A Retaliator behaves as a Hawk against Hawks, as a Dove against Doves and never escalates first. As in Section \ref{sec:2}, species share the same amount of resources $b>0$ and when escalating they have a $50\%$ chance of being injured. The payoff matrix then looks as follows:

$$R=\left( \begin{array}{ccccc}
\frac{b-c}{2}& &b& &\frac{b-c}{2}\\
\\
0& &\frac{b}{2}& &\frac{b}{2}\\
\\
\frac{b-c}{2}& &\frac{b}{2}& &\frac{b}{2}\\
\end{array} \right).$$

For instance, when $b=2$ and $c=4$:

$$R=\left( \begin{array}{rcccr}
-1& &2& &-1\\
0& &1& &1\\
-1& &1& &1\\
\end{array} \right).$$

This example was analyzed by Bomze \cite{Bomze1983} and the flow associated with the replicator equations for this game is shown in Figure 1\footnote{We use the Wolfram Mathematica project \cite{PhasePlane} in order to draw phase planes for this manuscript. }. Vertices of the triangle correspond to populations consisting of only Hawks, Doves or Retaliators. There are two fixed points and one pointwise-fixed line. The latter corresponds to Doves and Retaliators coexisting, at equilibrium, in all possible proportions. A pure strategy Hawk, represented by the point $[1,0,0]$, is a source. Any population with initial frequency of Hawks more than $0$ will evolve away from this point. A mixed strategy of Hawks and Doves, represented by $[\frac{1}{2},\frac{1}{2},0]$, is a sink. Hence, any population with an appropriately small portion of Retaliators  will tend to this equilibrium. However, if the number of Retaliators is sufficiently large, the mixture of Retaliators and Doves becomes evolutionary desirable and outcompetes the aggressive Hawks. This is a consequence of the high cost $c=4$.

\begin{figure}[h!]
\begin{center}
\includegraphics[scale=0.7]{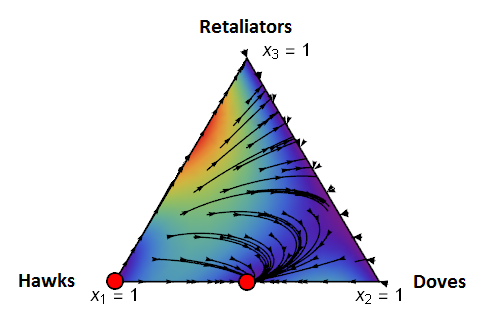}
\end{center}
\caption{The flow of the HDR-game}
\end{figure}

This game is a non-robust version of the HDR-game and any small perturbations of the entries of $R$ might change the behavior of the dynamics. We introduce incompetence in this case, for example, with the starting level $S$ as follows:
$$S=\left( \begin{array}{ccccc}
\frac{1}{2}& &\frac{1}{2}& &0\\
\\
\frac{1}{3}& &\frac{1}{3}& &\frac{1}{3}\\
\\
\frac{1}{4}& &\frac{1}{2}& &\frac{1}{4}\\
\end{array} \right).$$

Then the incompetence matrix for this example is
\begin{equation}
\label{HDRincompMatrix}
Q(\lambda)=\left( \begin{array}{ccccc}
\frac{1}{2}+\frac{1}{2}\lambda& &\frac{1}{2}-\frac{1}{2}\lambda& &0\\
\\
\frac{1}{3}-\frac{1}{3}\lambda& &\frac{1}{3}+\frac{2}{3}\lambda& &\frac{1}{3}-\frac{1}{3}\lambda\\
\\
\frac{1}{4}-\frac{1}{4}\lambda& &\frac{1}{2}-\frac{1}{2}\lambda& &\frac{1}{4}+\frac{3}{4}\lambda\\
\end{array} \right).
\end{equation}

Now it is easy to derive $\tilde{R}(\lambda)$ from (\ref{SimplifiedRM}) as

\begin{equation*}
\label{HDRRewardMatrix}
\tilde{R}(\lambda)=\left( \begin{array}{ccc}
0&\frac{3}{2}\lambda^2-\frac{1}{2}\lambda&-\frac{3}{2}\lambda^2-\frac{1}{2}\lambda\\
\\
\lambda^2+\frac{1}{6}\lambda+\frac{1}{6}&0&-\frac{1}{8}\lambda^2+\frac{1}{6}\lambda-\frac{1}{24}\\
\\
\frac{3}{8}\lambda^2-\frac{1}{4}\lambda-\frac{1}{8}&\frac{1}{2}\lambda^2-\frac{7}{12}\lambda+\frac{1}{12}&0\\
\end{array} \right).
\end{equation*}

Depending on the value of the incompetence parameter, $\lambda$, we obtain different qualitative behavior of the game dynamics. In particular, the set of critical points is $\Lambda=\{0,\frac{1}{7},\frac{1}{3},1\}$. These four critical values $\lambda^c$ and three intervals of $[0,1]$ induced by them determine seven regions with different qualitative behavior. The game flows for these regions can be found on Figure 2 and Figure 1 for $\lambda=1$. It is easy to verify that $\det(\tilde{R}(\lambda^c))=0$ for each $\lambda^c\in \Lambda$. Indeed, in this example $\Lambda=Z$ (see Definition 2).

\begin{figure}[h!]
\begin{center}
\includegraphics[scale=0.38]{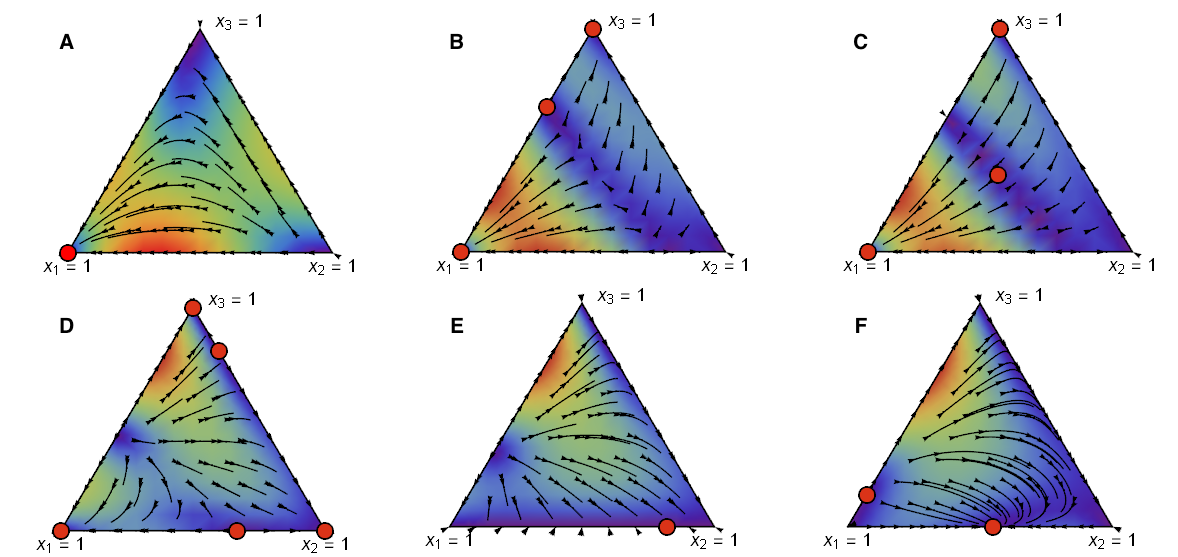}
\end{center}
\caption{The flow for the HDR-game for: A. $\lambda=0$, B. $\lambda=\frac{3}{25}$, C. $\lambda=\frac{1}{7}$, D. $\lambda=\frac{1}{4}$, E. $\lambda=\frac{1}{3}$, F. $\lambda=\frac{3}{5}$}
\end{figure}

Let us analyze changes in stability properties of the fixed points depending on the incompetence parameter. For sufficiently low level of competence we see that a Hawk strategy is preferable, especially for $\lambda\in [0,\frac{1}{7}]$. However, as species adapt to the environment and improve their competence, we observe that Retaliator and Dove strategies are competing with the aggressive Hawk behavior. That is, for the low competence, aggressive behavior is more preferable by natural selection. However, by adapting and improving their competence, species are more likely to choose less harmful strategies. In the region $\lambda\in (\frac{1}{3},1)$ we obtain a stabilized version of HDR game. Here Doves out-compete Retaliators: despite the fact that the Dove vertex is a saddle-point (see Panel (F) in Figure 2), when Retaliators are established they outcompete Hawks, and then Doves out-compete Retaliators. At $\lambda=1$ we observe an unstable game flow from Figure 1.

Let us now determine the existence of the ESS in this example. For the interval $\lambda\in [0,\frac{1}{3})$ a Hawk is the most preferable strategy, furthermore, it is an ESS. As $\lambda$ approaches $\frac{1}{3}$ the Hawk's competitive advantage weakens, and when it becomes a source after $\lambda=\frac{1}{3}$ (see Panel (E) and (F) in Figure 2), any trajectory runs away from this point. For $\lambda>\frac{1}{3}$ the probabilities of other incidental strategies tend to $0$, and a new ESS on the edge corresponding to Hawks-Doves starts to gain power. 

However, as previously mentioned, the evolution of fixed points depending on the incompetence parameter established above is a description of the evolution of a strategy choice. Of course, in nature, it is only possible to observe $\tilde{\mathbf{y}}(\lambda)$, namely, the current fixed point $\tilde{\mathbf{x}}(\lambda)$, which depends on the current level of incompetence and is randomized by the effect of the incompetence matrix $Q(\lambda)$ (see (\ref{execution})). That is according to (\ref{HDRincompMatrix}) for small $\lambda$ at the pure-Hawk-population ESS we may still observe Dove behavior with probability $\frac{1}{2}-\frac{1}{2}\lambda$. The probability of observing Retaliators is non-zero for larger $\lambda$ at the ESS on the Hawk-Dove edge (see left panel on Figure 3), where $\tilde{x}_1(\lambda),\;\tilde{x}_2(\lambda),\;\tilde{x}_3(\lambda)$ correspond to the red, blue and green curves, respectively. The plot of te corresponding $\tilde{y}_1(\lambda),\;\tilde{y}_2(\lambda),\;\tilde{y}_3(\lambda)$ probabilities can be found in Figure 3 (right panel), with the same color scheme.

\begin{figure}[h!]
\begin{center}
\includegraphics[scale=0.6]{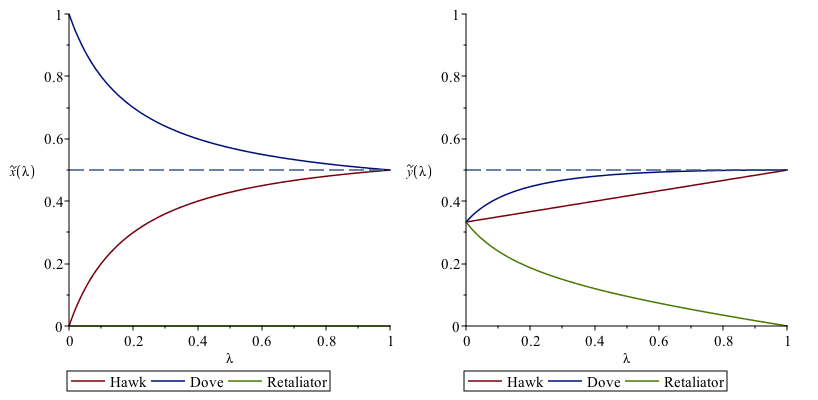}
\caption{The frequencies of HDR strategies at the Hawk-Dove edge fixed point (left panel) and probabilities of meeting HDR strategies at this point (right panel) depending on $\lambda$}
\end{center}
\end{figure}

\section{Conclusions}
\label{sec:concl}

When populations are unfamiliar with a new environment, it is a natural assumption that individuals may be prone to behavioral errors. Indeed, this has been observed within a diverse range of systems, from bacteria utilizing motility genes \cite{adler1966chemotaxis} to language learning for human beings \cite{Komarova2004,Komarova2001}. In this paper we proposed a generalized theoretical approach to this concept by incorporating a notion of incompetence into the central concept of evolutionary game theory: the replicator dynamics. 

Under incompetence, by analyzing replicator dynamics, we are analyzing the strategy choice. Hence, selection may be different to what occurs in the fully competent case. Moreover, the special structure of perturbations under incompetence contains hints as to where to look for qualitative changes in evolutionary dynamics. If dynamics are structurally stable, then incompetence must be reasonably high to affect the evolutionary stable outcome of the dynamics. However, structural stability is an elegant but rare condition for high-dimensional systems \cite{huttegger2010generic}. Hence, even small perturbations of the payoff matrix, corresponding to a small degree of incompetence, may affect the selection outcome in real-world systems. Additionally, in nature, we cannot ask species about their strategy choice. All we can do is to observe their strategy execution and, according to the incompetence matrix, whenever $\lambda<1$ we encounter stochasticity in the population behavior. Thus, if some behavioral types become extinct, they may still appear within the population as a manifestation of their mistakes. This may act as a redundancy, allowing species to utilize `lost' strategies if they become advantageous again, perhaps, due to changes in environmental conditions. That is, this implies a `memory' of extinct types may persist and lead to re-emergence of these types. It would be fascinating to design empirical experiments within suitable model systems, such as bacterial populations \cite{Frey2011,Lenski1991,Lenski1994,Lenski2000}, to test whether this phenomenon can be observed under laboratory conditions.

Here, we made the first step in the direction of introducing evolutionary games under incompetence. We considered only a one-parameter system with one population of species interacting. It will be natural to extend this to $n$-parametrized systems, where each strategy choice has its own adaptation parameter $\lambda_i$. That is, for the set of $n$ strategies, we could consider an incompetence vector $\lambda=(\lambda_1,...,\lambda_n)$, where $\lambda_i$ is the adaptation parameter for $i$-th behavioral type. Future work could also examine interactions between several populations, or applications of this theoretical approach within real-world systems.

\bibliographystyle{plain}
\bibliography{MyRefs}

\end{document}